\documentclass[10pt,a4paper,onecolumn]{article}
\usepackage{marginnote}
\usepackage{graphicx}
\usepackage{xcolor}
\usepackage{authblk,etoolbox}
\usepackage{titlesec}
\usepackage{calc}
\usepackage{tikz}
\usepackage{hyperref}
\usepackage{microtype}
\hypersetup{colorlinks,breaklinks,
            urlcolor=[rgb]{0.0, 0.5, 1.0},
            linkcolor=[rgb]{0.0, 0.5, 1.0}}
\usepackage{caption}
\usepackage{tcolorbox}
\usepackage{amssymb,amsmath}
\usepackage{ifxetex,ifluatex}
\usepackage{seqsplit}
\usepackage{xstring}

\usepackage{fixltx2e} 
\usepackage[
  backend=biber,
]{biblatex}
\bibliography{paper.bib}

\let\textttOrig=\texttt
\def\texttt#1{\expandafter\textttOrig{\seqsplit{#1}}}
\renewcommand{\seqinsert}{\ifmmode
  \allowbreak
  \else\penalty6000\hspace{0pt plus 0.02em}\fi}

\makeatletter
\let\href@Orig=\href
\def\href@Urllike#1#2{\href@Orig{#1}{\begingroup
    \def\Url@String{#2}\Url@FormatString
    \endgroup}}
\def\href@Notdoi#1#2{\def\tempa{#1}\def\tempb{#2}%
  \ifx\tempa\tempb\relax\href@Urllike{#1}{#2}\else
  \href@Orig{#1}{#2}\fi}
\def\href#1#2{%
  \IfBeginWith{#1}{https://doi.org}%
  {\href@Urllike{#1}{#2}}{\href@Notdoi{#1}{#2}}}
\makeatother

\usepackage[top=3.2cm, bottom=2.9cm, right=1.5cm, left=1.0cm,
            headheight=1.9cm, reversemp, includemp, marginparwidth=3.9cm]{geometry}

\titleformat{\section}
  {\normalfont\sffamily\Large\bfseries}
  {}{0pt}{}
\titleformat{\subsection}
  {\normalfont\sffamily\large\bfseries}
  {}{0pt}{}
\titleformat{\subsubsection}
  {\normalfont\sffamily\bfseries}
  {}{0pt}{}
\titleformat*{\paragraph}
  {\sffamily\normalsize}

\usepackage{fancyhdr}
\makeatletter
\let\ps@plain\ps@fancy
\fancyheadoffset[L]{4.5cm}
\fancyfootoffset[L]{4.5cm}

\definecolor{linky}{rgb}{0.0, 0.5, 1.0}

\newtcolorbox{repobox}
   {colback=red, colframe=red!75!black,
     boxrule=0.5pt, arc=2pt, left=6pt, right=6pt, top=3pt, bottom=3pt}

\newcommand{\ExternalLink}{%
   \tikz[x=1.2ex, y=1.2ex, baseline=-0.05ex]{%
       \begin{scope}[x=1ex, y=1ex]
           \clip (-0.1,-0.1)
               --++ (-0, 1.2)
               --++ (0.6, 0)
               --++ (0, -0.6)
               --++ (0.6, 0)
               --++ (0, -1);
           \path[draw,
               line width = 0.5,
               rounded corners=0.5]
               (0,0) rectangle (1,1);
       \end{scope}
       \path[draw, line width = 0.5] (0.5, 0.5)
           -- (1, 1);
       \path[draw, line width = 0.5] (0.6, 1)
           -- (1, 1) -- (1, 0.6);
       }
   }

\patchcmd{\@maketitle}{center}{flushleft}{}{}
\patchcmd{\@maketitle}{center}{flushleft}{}{}
\patchcmd{\@maketitle}{\LARGE}{\LARGE\sffamily}{}{}
\def\maketitle{{%
  
  \AB@maketitle}}
\makeatletter
\renewcommand\AB@affilsepx{ \protect\Affilfont}
\renewcommand\AB@affilnote[1]{{\bfseries #1}\hspace{3pt}}
\renewcommand{\affil}[2][]%
   {\newaffiltrue\let\AB@blk@and\AB@pand
      \if\relax#1\relax\def\AB@note{\AB@thenote}\else\def\AB@note{#1}%
        \setcounter{Maxaffil}{0}\fi
        \begingroup
        \let\href=\href@Orig
        \let\texttt=\textttOrig
        \let\protect\@unexpandable@protect
        \def\thanks{\protect\thanks}\def\footnote{\protect\footnote}%
        \@temptokena=\expandafter{\AB@authors}%
        {\def\\{\protect\\\protect\Affilfont}\xdef\AB@temp{#2}}%
         \xdef\AB@authors{\the\@temptokena\AB@las\AB@au@str
         \protect\\[\affilsep]\protect\Affilfont\AB@temp}%
         \gdef\AB@las{}\gdef\AB@au@str{}%
        {\def\\{, \ignorespaces}\xdef\AB@temp{#2}}%
        \@temptokena=\expandafter{\AB@affillist}%
        \xdef\AB@affillist{\the\@temptokena \AB@affilsep
          \AB@affilnote{\AB@note}\protect\Affilfont\AB@temp}%
      \endgroup
       \let\AB@affilsep\AB@affilsepx
}
\makeatother

\renewcommand\Affilfont{\sffamily\small\mdseries}
\ifnum 0\ifxetex 1\fi\ifluatex 1\fi=0 
  \usepackage[T1]{fontenc}
  \usepackage[utf8]{inputenc}

\else 
  \ifxetex
    \usepackage{mathspec}
  \else
    \usepackage{fontspec}
  \fi
  \defaultfontfeatures{Ligatures=TeX,Scale=MatchLowercase}

\fi
\IfFileExists{upquote.sty}{\usepackage{upquote}}{}
\IfFileExists{microtype.sty}{%
\usepackage{microtype}
\UseMicrotypeSet[protrusion]{basicmath} 
}{}

\usepackage{hyperref}
\hypersetup{unicode=true,
            pdftitle={qnm: A Python package for calculating Kerr quasinormal modes, separation constants, and spherical-spheroidal mixing coefficients},
            pdfauthor={Stein},
            pdfborder={0 0 0},
            breaklinks=true}
\let\addcontentslineOrig=\addcontentsline
\def\addcontentsline#1#2#3{\bgroup
  \let\texttt=\textttOrig\addcontentslineOrig{#1}{#2}{#3}\egroup}
\let\markbothOrig\markboth
\def\markboth#1#2{\bgroup
  \let\texttt=\textttOrig\markbothOrig{#1}{#2}\egroup}
\let\markrightOrig\markright
\def\markright#1{\bgroup
  \let\texttt=\textttOrig\markrightOrig{#1}\egroup}

\usepackage{graphicx,grffile}
\makeatletter
\def\maxwidth{\ifdim\Gin@nat@width>\linewidth\linewidth\else\Gin@nat@width\fi}
\def\maxheight{\ifdim\Gin@nat@height>\textheight\textheight\else\Gin@nat@height\fi}
\makeatother
\setkeys{Gin}{width=\maxwidth,height=\maxheight,keepaspectratio}
\IfFileExists{parskip.sty}{%
\usepackage{parskip}
}{
\setlength{\parindent}{0pt}
\setlength{\parskip}{6pt plus 2pt minus 1pt}
}
\ifx\paragraph\undefined\else
\let\oldparagraph\paragraph
\renewcommand{\paragraph}[1]{\oldparagraph{#1}\mbox{}}
\fi
\ifx\subparagraph\undefined\else
\let\oldsubparagraph\subparagraph
\renewcommand{\subparagraph}[1]{\oldsubparagraph{#1}\mbox{}}
\fi

\title{qnm: A Python package for calculating Kerr quasinormal modes, separation
constants, and spherical-spheroidal mixing coefficients}

        \author[1]{Leo C. Stein}
    
      \affil[1]{Department of Physics and Astronomy, The University of Mississippi,
University, MS 38677, USA}
  \date{\vspace{-5ex}}

\begin{document}
\maketitle

\marginpar{
  \sffamily\small

  {\bfseries DOI:} \href{https://doi.org/10.21105/joss.01683}{\color{linky}{10.21105/joss.01683}}

  \vspace{2mm}

  {\bfseries Software}
  \begin{itemize}
    \setlength\itemsep{0em}
    \item \href{https://github.com/openjournals/joss-reviews/issues/1683}{\color{linky}{Review}} \ExternalLink
    \item \href{https://github.com/duetosymmetry/qnm}{\color{linky}{Repository}} \ExternalLink
    \item \href{https://doi.org/10.5281/zenodo.3459790}{\color{linky}{Archive}} \ExternalLink
  \end{itemize}

  \vspace{2mm}

  {\bfseries Submitted:} 21 August 2019\\
  {\bfseries Published:} 01 October 2019

  \vspace{2mm}
  {\bfseries License}\\
  \hyphenpenalty=1000
  Authors of papers retain copyright and release the work under a Creative Commons Attribution 4.0 International License (\href{http://creativecommons.org/licenses/by/4.0/}{\color{linky}{CC-BY}}).
}
\vspace{-2mm}
\hypertarget{background}{%
\section{Background}\label{background}}
\vspace{-2mm}

Black holes can be characterized from far away by their spectroscopic
gravitational-wave ``fingerprints,'' in analogy to electromagnetic
spectroscopy of atoms, ions, and molecules. The idea of using the
quasi-normal modes (QNMs) of black holes (BHs) for gravitational-wave
(GW) spectroscopy was first made explicit by Detweiler (1980). QNMs of
rotating Kerr BHs in general relativity (GR) depend only on the mass and
spin of the BH. Thus GWs containing QNMs can be used to infer the
remnant BH properties in a binary merger, or as a test of GR by checking
the consistency between the inspiral and ringdown portions of a GW
signal (Abbott and others 2016; Isi et al. 2019).

For a review of QNMs see Berti, Cardoso, and Starinets (2009). A Kerr
BH's QNMs are the homogeneous (source-free) solutions to the Teukolsky
equation (Teukolsky 1973) subject to certain physical conditions. The
Teukolsky equation can apply to different physical fields based on their
spin-weight \(s\); for gravitational perturbations, we are interested in
\(s=-2\) (describing the Newman-Penrose scalar \(\psi_4\)). The physical
conditions for a QNM are quasi-periodicity in time, of the form
\(\propto e^{-i \omega t}\) with complex \(\omega\); conditions of
regularity, and that the solution has waves that are only going down the
horizon and out at spatial infinity. Separating the radial/angular
Teukolsky equations and imposing these conditions gives an eigenvalue
problem where the frequency \(\omega\) and separation constant \(A\)
must be found simultaneously. This eigenvalue problem has a countably
infinite, discrete spectrum labeled by angular harmonic numbers
\((\ell, m)\) with \(\ell\ge 2\) (or \(\ell \ge |s|\) for fields of
other spin weight), \(-\ell \le m \le +\ell\), and overtone number
\(n \ge 0\).

There are several analytic techniques, e.g.~one presented by Dolan and
Ottewill (2009), to approximate the desired complex frequency and
separation constant \((\omega_{\ell, m, n}(a), A_{\ell, m, n}(a))\) as a
function of spin parameter \(0 \le a < M\) (we follow the convention of
using units where the total mass is \(M=1\)). These analytic techniques
are useful as starting guesses before applying the numerical method of
Leaver (1985) for root-polishing. Leaver's method uses Frobenius
expansions of the radial and angular Teukolsky equations to find 3-term
recurrence relations that must be satisfied at a complex frequency
\(\omega\) and separation constant \(A\). The recurrence relations are
made numerically stable to find so-called minimal solutions by being
turned into infinite continued fractions. In Leaver's approach, there
are thus two ``error'' functions \(E_r(\omega, A)\) and
\(E_a(\omega, A)\) (each depending on \(a, \ell, m, n\)) which are given
as infinite continued fractions, and the goal is to find a pair of
complex numbers \((\omega, A)\) which are simultaneous roots of both
functions. This is typically accomplished by complex root-polishing,
alternating between the radial and angular continued fractions.

A refinement of this method was put forth by Cook and Zalutskiy (2014)
(see also Appendix A of Hughes (2000)). Instead of solving the angular
Teukolsky equation ``from the endpoint'' using Leaver's approach, one
can use a spectral expansion with a good choice of basis functions. The
solutions to the angular problem are the \emph{spin-weighted spheroidal
harmonics}, and the appropriate spectral basis are the spin-weighted
\emph{spherical} harmonics. This expansion is written as (spheroidal on
the left, sphericals on the right):

\[{}_s Y_{\ell m}(\theta, \phi; a\omega) = {\sum_{\ell'=\ell_{\min} (s,m)}^{\ell_{\max}}} C_{\ell' \ell m}(a\omega)\ {}_s Y_{\ell' m}(\theta, \phi) \,,\]

where \(\ell_{\min} = \max(|m|, |s|)\), and the coefficients
\(C_{\ell' \ell m}(a\omega)\) are called the spherical-spheroidal mixing
coefficients (we follow the conventions of Cook and Zalutskiy (2014),
but compare Berti and Klein (2014)). When recast in this spectral form,
the angular equation becomes very easy to solve via standard matrix
eigenvector routines, see Cook and Zalutskiy (2014) for details. If one
picks values for \((s, \ell, m, a, \omega)\), then the separation
constant \(A(a\omega)\) is returned as an eigenvalue, and a vector of
mixing coefficients \(C_{\ell' \ell m}(a\omega)\) are returned as an
eigenvector. From this new point of view there is now only one error
function to root-polish, \(E_r(\omega) = E_r(\omega, A(\omega))\) where
the angular separation constant is found from the matrix method at any
value of \(\omega\). Polishing roots of \(E_r\) proceeds via any
standard 2-dimensional root-finding or optimization method.

The main advantage of the spectral approach is rapid convergence, and
getting the spherical-spheroidal mixing coefficients ``for free'' since
they are found in the process of solving the spectral angular eigenvalue
problem.

\vspace{-2mm}
\hypertarget{summary}{%
\section{Summary}\label{summary}}
\vspace{-2mm}

\texttt{qnm} is an open-source Python package for computing the Kerr QNM
frequencies, angular separation constants, and spherical-spheroidal
mixing coefficients, for given values of \((\ell, m, n)\) and spin
\(a\). There are several QNM codes available, but some (London 2017)
implement either analytic fitting formulae (which only exist for a range
of \(s, \ell, m, n\)) or interpolation from tabulated data (so the user
can not root-polish); others (Berti 2010) are in proprietary languages
such as Mathematica. We are not aware of any packages that provide
spherical-spheroidal mixing coefficients, which are necessary for
multi-mode ringdown GW modeling.

The \texttt{qnm} package includes a Leaver solver with the
Cook-Zalutskiy spectral approach to the angular sector, thus providing
mixing coefficients. We also include a caching mechanism to avoid
repeating calculations. When the user wants to solve at a new value of
\(a\), the cached data is used to interpolate a good initial guess for
root-polishing. We provide a large cache of low \(\ell, m, n\) modes so
the user can start interpolating right away, and this precomputed cache
can be downloaded and installed with a single function call. We have
adapted the core algorithms so that \texttt{numba} (Lam, Pitrou, and
Seibert 2015) can just-in-time compile them to optimized, machine-speed
code. We rely on \texttt{numpy} (Walt, Colbert, and Varoquaux 2011) for
common operations such as solving the angular eigenvalue problem, and we
rely on \texttt{scipy} (Jones et al. 2001) for two-dimensional
root-polishing, and interpolating from the cache before root-polishing.

This package should enable researchers to perform ringdown modeling of
gravitational-wave data in Python, without having to interpolate into
precomputed tables or write their own Leaver solver. The author and
collaborators are already using this package for multiple active
research projects. By creating a self-documented, open-source code, we
hope to alleviate the high frequency of re-implemenation of Leaver's
method, and instead focus efforts on making a single robust, fast,
high-precision, and easy-to-use code for the whole community. In the
future, this code can be extended to incorporate new features (like
special handling of algebraically special modes) or to apply to more
general BH solutions (e.g., solving for QNMs of Kerr-Newman or Kerr-de
Sitter).

Development of \texttt{qnm} is hosted on
\href{https://github.com/duetosymmetry/qnm}{GitHub} and distributed
through \href{https://pypi.org/project/qnm/}{PyPI}; it can be installed
with the single command \texttt{pip\ install\ qnm}. Documentation is
automatically built on \href{https://qnm.readthedocs.io/}{Read the
Docs}, and can be accessed interactively via Python docstrings.
Automated testing is run on
\href{https://travis-ci.org/duetosymmetry/qnm}{Travis CI}. The
\texttt{qnm} package is part of the \href{https://bhptoolkit.org/}{Black
Hole Perturbation Theory Toolkit}.

\vspace{-2mm}
\hypertarget{acknowledgements}{%
\section{Acknowledgements}\label{acknowledgements}}
\vspace{-2mm}

We acknowledge E Berti and GB Cook for helpful correspondence and for
making testing data available. We further acknowledge M Giesler, I
Hawke, Duncan Macleod, L Magaña Zertuche, Matt Pitkin, and V Varma for
contributions/testing/feedback/suggestions.

\vspace{-2mm}
\hypertarget{references}{%
\section*{References}\label{references}}
\addcontentsline{toc}{section}{References}
\vspace{-2mm}

\hypertarget{refs}{}
\leavevmode\hypertarget{ref-TheLIGOScientific:2016src}{}%
Abbott, B. P., and others. 2016. ``Tests of general relativity with
GW150914.'' \emph{Phys. Rev. Lett.} 116 (22): 221101.
\url{https://doi.org/10.1103/PhysRevLett.116.221101}.

\leavevmode\hypertarget{ref-BertiWeb}{}%
Berti, Emanuele. 2010. ``Berti's ringdown web page.''
\url{https://pages.jh.edu/~eberti2/ringdown/}.

\leavevmode\hypertarget{ref-Berti:2009kk}{}%
Berti, Emanuele, Vitor Cardoso, and Andrei O. Starinets. 2009.
``Quasinormal modes of black holes and black branes.'' \emph{Class.
Quant. Grav.} 26: 163001.
\url{https://doi.org/10.1088/0264-9381/26/16/163001}.

\leavevmode\hypertarget{ref-Berti:2014fga}{}%
Berti, Emanuele, and Antoine Klein. 2014. ``Mixing of spherical and
spheroidal modes in perturbed Kerr black holes.'' \emph{Phys. Rev.} D90
(6): 064012. \url{https://doi.org/10.1103/PhysRevD.90.064012}.

\leavevmode\hypertarget{ref-Cook:2014cta}{}%
Cook, Gregory B., and Maxim Zalutskiy. 2014. ``Gravitational
perturbations of the Kerr geometry: High-accuracy study.'' \emph{Phys.
Rev.} D90 (12): 124021.
\url{https://doi.org/10.1103/PhysRevD.90.124021}.

\leavevmode\hypertarget{ref-Detweiler:1980gk}{}%
Detweiler, Steven L. 1980. ``Black holes and gravitational waves. III.
The resonant frequencies of rotating holes.'' \emph{Astrophys. J.} 239:
292--95. \url{https://doi.org/10.1086/158109}.

\leavevmode\hypertarget{ref-Dolan:2009nk}{}%
Dolan, Sam R., and Adrian C. Ottewill. 2009. ``On an Expansion Method
for Black Hole Quasinormal Modes and Regge Poles.'' \emph{Class. Quant.
Grav.} 26: 225003. \url{https://doi.org/10.1088/0264-9381/26/22/225003}.

\leavevmode\hypertarget{ref-Hughes:1999bq}{}%
Hughes, Scott A. 2000. ``The Evolution of circular, nonequatorial orbits
of Kerr black holes due to gravitational wave emission.'' \emph{Phys.
Rev.} D61 (8): 084004. \url{https://doi.org/10.1103/PhysRevD.61.084004}.

\leavevmode\hypertarget{ref-Isi:2019aib}{}%
Isi, Maximiliano, Matthew Giesler, Will M. Farr, Mark A. Scheel, and
Saul A. Teukolsky. 2019. ``Testing the no-hair theorem with GW150914.''
\emph{Phys. Rev. Lett.} 123 (11): 111102.
\url{https://doi.org/10.1103/PhysRevLett.123.111102}.

\leavevmode\hypertarget{ref-SciPy}{}%
Jones, Eric, Travis Oliphant, Pearu Peterson, and others. 2001. ``SciPy:
Open Source Scientific Tools for Python.'' \url{http://www.scipy.org}.

\leavevmode\hypertarget{ref-Numba}{}%
Lam, Siu Kwan, Antoine Pitrou, and Stanley Seibert. 2015. ``Numba: A
LLVM-based Python JIT Compiler.'' In \emph{Proceedings of the Second
Workshop on the Llvm Compiler Infrastructure in Hpc}, 7:1--7:6. LLVM
'15. New York, NY, USA: ACM.
\url{https://doi.org/10.1145/2833157.2833162}.

\leavevmode\hypertarget{ref-Leaver:1985ax}{}%
Leaver, E. W. 1985. ``An Analytic representation for the quasi normal
modes of Kerr black holes.'' \emph{Proc. Roy. Soc. Lond.} A402: 285--98.
\url{https://doi.org/10.1098/rspa.1985.0119}.

\leavevmode\hypertarget{ref-LondonGithub}{}%
London, Lionel. 2017. ``London's QNM codes @ 455f2d5.''
\url{https://github.com/llondon6/kerr_public}.

\leavevmode\hypertarget{ref-Teukolsky:1973ha}{}%
Teukolsky, Saul A. 1973. ``Perturbations of a rotating black hole. 1.
Fundamental equations for gravitational electromagnetic and neutrino
field perturbations.'' \emph{Astrophys. J.} 185: 635--47.
\url{https://doi.org/10.1086/152444}.

\leavevmode\hypertarget{ref-NumPy}{}%
Walt, Stéfan van der, S. Chris Colbert, and Gaël Varoquaux. 2011. ``The
NumPy Array: A Structure for Efficient Numerical Computation.''
\emph{Computing in Science \& Engineering} 13 (2): 22--30.
\url{https://doi.org/10.1109/MCSE.2011.37}.

\end{document}